\documentclass{article}

\usepackage{bm, amsmath, amssymb, url, hyperref}
\usepackage{pdflscape, longtable}
\usepackage{tabularx}
\usepackage[singlelinecheck=false, width=\textwidth]{caption}
\usepackage{multirow}
\usepackage{rotating}
\usepackage{dcolumn}
\usepackage{float}
\usepackage{Sweave}

\usepackage{natbib}
\usepackage{appendix}

\hypersetup{
  colorlinks = true,
  urlcolor = blue,  
  linkcolor = blue,
}

\begin{document}

\title{Computing a consensus journal meta-ranking using paired comparisons and adaptive lasso estimators} 

\author{Laura Vana \and Ronald Hochreiter \and Kurt Hornik}

\maketitle

\begin{abstract} 
In a ``publish-or-perish culture'', the ranking of scientific journals plays a central role in assessing performance in the current research environment. With a wide range of existing methods and approaches to deriving journal rankings, meta-rankings have gained popularity as a means of aggregating different information sources. In this paper, we propose a method to create a consensus meta-ranking using heterogeneous journal rankings. Using a parametric model for paired comparison data we estimate quality scores for 58 journals in the OR/MS community, which together with a shrinkage procedure allows for the identification of clusters of journals with similar quality. The use of paired comparisons provides a flexible framework for deriving a consensus score while eliminating the problem of data missingness.
\end{abstract}


\sloppy
\section{Introduction}

While the so-called ``publish-or-perish culture'' has been widely discussed and criticized mainly because of intense publication pressure \citep[see for example][]{adler2009knowledge, frey2010withering, willmott2011journal}, an increasingly competitive research environment is in need of  performance metrics. Generally, the reputation and quality of the research outlets in which scholars publish their work together with received citations in peer-reviewed academic journals remain the main criteria for assessing research quality. The main research outlets and means of knowledge dissemination in each business discipline are the academic journals in which its research is published \citep{meredith2011knowledge, fry2013outlets}. In an attempt to reflect the impact and quality of journals, rankings therefore lie at the core of research assessment. Whether it is used by universities for hiring or promotion/tenure purposes, by editors to underline the importance of their journals or by publishers who aim at maximizing their revenue, a ranking supports decision making. Especially due to this importance in academic life the number of applied methods to create rankings is increasing continuously. 

While there are many different rankings available, it is often hard to pick the most appropriate ranking for a special purpose or a certain institution. Different stakeholders often prefer a different ranking for various reasons, such that the creation of meta-rankings is the only way to comfort all involved parties. Moreover, the disadvantages of the conventional approaches to creating journal rankings, with citation-based and survey-based approaches being the most popular, have been outlined in the literature \citep[i.e.,][]{serenko2011comparing}. In this light, meta-rankings have gained popularity due to their the advantage of building on previous work and aggregating the available information rather than building against it. 

In this paper, we propose a novel method to aggregate heterogeneous rankings using adaptive lasso estimators and apply the method to obtain a consensus meta-ranking for 58 established journals in the Operations Research/Management Science (OR/MS) community. 

The parametric Bradley-Terry model \citep{bradley1952rank} for paired comparison data is employed for estimating journal quality scores as an alternative to the popular data envelopment analysis (DEA) method \citep[used in i.e.,][]{fry2013outlets, tuselmann2015towards}. 

The use of paired comparisons provides a flexible framework for deriving a consensus score using different sources, with no strict restriction on the number of journals and the number of rankings used in the analysis. Together with a shrinkage procedure, the proposed approach allows the identification of clusters of journals with similar quality \citep{masarotto2012ranking}. The variability of the parameter estimators is determined using the parametric bootstrap.

We also investigate whether the consensus meta-ranking changes significantly when using only information from rankings published in 2013 and find that results remain mainly stable, with few significant changes.

\cite{tuselmann2015towards} discuss several shortcomings of meta-studies, among which arbitrary inclusion or datedness of journal rankings and journals, inadequate treatment of missing data and treatment of ordinal data as metric. 
We address these shortcomings in the following way: 
(i) we base our analysis on an extensive list of 31 journal rankings that are used by stakeholders in decision making. An overview of the rankings is provided in Section~\ref{sect:journalsandrankings} and our descriptive statistics show that the significant correlations at a 5\% significance level are positive. However, each of the rankings contains valuable information because, while having the common goal of assessing journal quality, they proceed differently in defining comparison measures. (ii) Our approach eliminates the evident problem of missingness in journal lists, while existing data need not be disregarded due to the use of pairwise comparisons between journals as expressed by the various ranking lists \citep[similar to][]{cook2010aggregating, theussl2010can}. (iii) The ordinal nature of rankings is kept by the pairwise comparison data. In addition, the parametric model is adapted to appropriately handle ties.

This paper is organized as follows. In Section~\ref{sect:lit} we present an overview of existing ranking approaches. Section~\ref{sect:journalsandrankings} presents the journals and journal ranking lists used in the study, Section~\ref{sect:model} introduces the methodology used for creating a consensus journal rating. The results are discussed in Section~\ref{sect:results} and Section~\ref{sect:concl} concludes.
 
\section{Ranking approaches}\label{sect:lit}
Even if quality is abstract by nature, several approaches have been proposed for assessing journal quality. Survey-based rankings are primarily created by universities or associations, who ask scholars with different affiliations to rate journals on a usually ordinal scale. Academic research relying on survey-based quality assessment of Operations Research, Management Science, Production \& Operations Management (OR/MS/POM) journals has been conducted by \cite{barman1991empirical, barman2001perceived} or \cite{olson2005top}, who surveyed faculty members of 25 business schools in the US in 2000 and 2002.

Citation-based metrics are mainly published by commercial providers like Thomson Reuters or Elsevier. The first academic studies to conduct a citation-based analysis of journals in the Operations Management (OM) field are \cite{vokurka1996} and \cite{goh1996empirical}. A recent citation-based analysis focused on the OR/MS field is 
\cite{xu2011evaluating}, who use Google's PageRank method to rank 31 OR/MS journals. The PageRank method is citation-based and differentiates citations by quality/source (i.e., citations from higher impact journals should outweigh citations from lower impact journals). This method is in essence similar to the \emph{Eigenfactor\textsuperscript{\textregistered} Score}, which will be discussed briefly in Section~\ref{sect:journalsandrankings}. 
Hybrid rankings are a combination of the subjective and objective approach \citep[i.e.,][]{zhou2001journal}, while author-based approaches take into consideration also the author affiliation.    
\cite{holsapple2010behavior} perform a behaviour-based analysis, by examining the publishing behaviors of
tenured OM researchers at leading research universities in the US.
 Extensive comparisons among the different methods as well as shortcomings can be found in i.e., \cite{donohue2000multi}, \cite{frey2010rankings}, \cite{serenko2011comparing}.  
   
 Considering the setbacks of the approaches mentioned above, the meta-ranking approach arises naturally as it tries to reconcile the variety of methods by using available information to build a composite journal ranking \citep{cook2010aggregating}. In the OM field several meta-analyses have combined information from prior academic studies. \cite{petersen2011journal} provide the first meta-analysis to examine the ranking of journals in the OM field by using 5 prior studies.
\cite{meredith2011knowledge} build journal rankings using official in-house journal lists of AACSB-accredited business schools and compare their results with 12 ranking studies published during 1990 to 2009.
\cite{fry2013outlets} use the information from 15 previous OM journal ranking studies and provide a meta-analysis through DEA for assessing journal quality.
A recent study is provided by \cite{tuselmann2015towards}, who base their study on 10 rankings from Harzing's JQL data and the \emph{Impact Factor} and compute an aggregate rating of journals in different subject areas by using DEA and random forests, with particular emphasis on OR/MS/POM. 
\section{Journals and quality assessment}
\label{sect:journalsandrankings}

We select 58 journals established in the OR/MS community which are also part of Thomson Reuters' Journal Citation Reports \citep{jcr}. The complete list of journals can be found in Table~\ref{tab:journals}. 

We collect an extensive list of 31 rankings still in use in 2013, both citation and survey-based. We match the rankings to the journal list using ISSN codes.
From Thomson Reuters' Journal Citation Reports we obtain citation-based journal metrics like the \emph{Impact Factor}, \emph{5-Year Impact Factor}, \emph{Immediacy Index} and \emph{Cited Half-Life}. The \emph{Impact Factor} is one of the best known indexes and measures the number of citations received by a journal in the two preceding years normalized by the number of citable items in that journal. The \emph{5-Year Impact Factor} extends the time period used for calculation to five years. The \emph{Immediacy Index} measures the number of citations received by a journal in the same year, while \emph{Cited Half-Life} measures the median age of a journal's cited articles in a certain year.
Other citation metrics freely available are the \emph{Eigenfactor\textsuperscript{\textregistered} Score} and its normalized version the \emph{Article Influence\textsuperscript{\textregistered} Score}. These metrics assign different weights to different sources of citations received by a journal and exclude self-citations. The time period used for calculation is five years\footnote{See more on the methodology at \url{http://www.eigenfactor.org/methods}.}.

From \url{www.journalmetrics.com}, the \emph{Source Normalized Impact per Paper} metric (SNIP) corrects for differences between fields by dividing the number of citations per paper in a journal by the citable items in the journal's subject field. The \emph{Impact per Publication} (IPP) is a metric similar to Thomson Reuters' \emph{Impact Factor} and measures the average citations to peer-reviewed papers published in the three previous years. A third metric is the \emph{SCImago Journal Rank} (SJR) which is in essence similar to the \emph{Eigenfactor\textsuperscript{\textregistered} Score}. Detailed methodology can be found on the website.

An important source of journal rankings (mostly survey-based) is the Harzing Journal Quality List (JQL). The 52nd edition of the JQL data \citep{jql} contains 22 rankings provided by different (research) institutions or scientific studies. Most of the JQL rankings are on an ordinal scale (see Table~\ref{tab:ranking}). 

All rankings used in the analysis are listed in Table~\ref{tab:ranking}. Out of 31 rankings, 13 were compiled in 2013. The other 18 are dated in the range from 2001 to 2012, but are still used in different contexts for assessing journal quality.

Due to the (mostly) ordinal nature of the rankings, the presence of ties and incomplete journal lists, we assess the degree of association between the different sources by using the rank correlation coefficient proposed by \cite{emond2002new}, which corresponds to the unique association measure of \cite{kemeny1962preference} and satisfies four basic axioms that should apply to any distance measure between two weak orderings. The metric is denoted by $\tau_x$ and is an extension of Kendall's $\tau$ which handles ties more appropriately  \citep[see also][]{hornik2006validation}:
\[\tau_x^{(k_1, k_2)} = 1 - \frac{\sum_{i=1}^{N_c}\sum_{j=1}^{N_c}|a_{ijk_1} - a_{ijk_2}|}{N_c(N_c-1)},\]
where $N_c$ is the number of journals rated in at least one of the rankings $R_{k_1}$ and $R_{k_2}$ and $a_{ijk}$ is defined as:
\begin{displaymath}
a_{ijk} =  \begin{cases}
1 & \mbox{if ranking $R_{k}$ rates journal $J_{i}$ higher than or tied with as journal $J_{j}$},\\
0 & \mbox{if $i = j$ or if $R_{k}$ does not rate $J_{i}$ or/and $J_{j}$},\\
-1 & \mbox{otherwise}.\\
\end{cases}
\end{displaymath}

We present the $\tau_x$ correlation coefficients in Table~\ref{tab:rstats}, together with the number of journals rated in the sample for each ranking. For each rank correlation coefficient, we compute $p$-values using 1000 bootstrap samples. Column ``MLE'' presents the correlation of the ranking produced using the method described in Section~\ref{sect:model} with the other 31 rankings. 
The citation-based indexes of Thomson Reuters, the \emph{Eigenfactor\textsuperscript{\textregistered} Score}, the \emph{Article Influence\textsuperscript{\textregistered} Score}  and the Elsevier metrics are mostly significantly correlated. Not surprisingly, the metric that is not significantly correlated with the rest is \emph{Cited Half-Life}, as it is the only metric based on the age of cited articles. We observe that at 5\% significance level the significant rank correlations are positive, but the degree of the association varies, indicating that the quality assessment of journals differs across different metrics or sources. 

We have in total $16452$ pairwise comparisons between the 58 journals and each pair of journals has been on average compared $10$ times.

\begin{sidewaystable}[ph!]
  \vspace{2cm}
\captionsetup{width=\textheight}
\caption[.]{Number of rated journals ($N_R$) and $\tau_x$ metric; $p$-values based on 1000 bootstrap samples.}
\begin{tiny}  
  \begin{tabular}{lc D{.}{.}{0.6} D{.}{.}{0.6}D{.}{.}{0.6} D{.}{.}{0.6} D{.}{.}{0.6}
    D{.}{.}{0.6} D{.}{.}{0.6}D{.}{.}{0.6} D{.}{.}{0.6} D{.}{.}{0.6}
    D{.}{.}{0.6} D{.}{.}{0.6}D{.}{.}{0.6} D{.}{.}{0.6} D{.}{.}{0.6}D{.}{.}{0.6}}
\hline
& $N_R$ &\multicolumn{1}{l}{IF5Y} &\multicolumn{1}{l}{IMMI} & \multicolumn{1}{l}{CHL} & \multicolumn{1}{l}{EFS} & \multicolumn{1}{l}{AIS} & 
\multicolumn{1}{l}{SNIP} & \multicolumn{1}{l}{IPP} & \multicolumn{1}{l}{SJR} & \multicolumn{1}{l}{Wie01} & \multicolumn{1}{l}{Vhb03} & \multicolumn{1}{l}{Bjm04} & \multicolumn{1}{l}{Hkb05} & \multicolumn{1}{l}{Theo05} & \multicolumn{1}{l}{Ejis07} & \multicolumn{1}{l}{EjisCI07} & \multicolumn{1}{l}{UQ07}\\ 

  \hline
IF & 58 & 0.74^{***} & 0.51^{***} & -0.04    & 0.48^{***} & 0.42^{***} & 0.09^{***} & 0.11^{***} & 0.08^{**}  & 0.02    & 0.02    & -0.02    &  0.00    &  0.00    & 0.05    & 0.07^{*}   & 0.05^{*}   \\ 
  IF5Y & 56 &  & 0.45^{***} & -0.02    & 0.48^{***} & 0.56^{***} & 0.13^{***} & 0.14^{***} & 0.12^{***} & 0.03    & 0.05    &  0.02    &  0.00    &  0.00    & 0.06    & 0.07    & 0.05    \\ 
  IMMI & 56 &  &  &  0.02    & 0.36^{***} & 0.31^{***} & 0.06    & 0.08^{*}   & 0.08^{*}   & 0.03    & 0.03    &  0.01    & -0.01    &  0.00    & 0.06    & 0.06    & 0.02    \\ 
  CHL & 57 &  &  &  & 0.17    & 0.12    & 0.00    & 0.02    & 0.02    & 0.06^{*}   & 0.04    &  0.03    &  0.04    &  0.00    & 0.04    & 0.05    & 0.02    \\ 
  EFS & 58 &  &  &  &  & 0.51^{***} & 0.06^{*}   & 0.06^{*}   & 0.08^{**}  & 0.03    & 0.05    &  0.00    &  0.01    &  0.00    & 0.09^{**}  & 0.12^{**}  & 0.04^{*}   \\ 
  AIS & 56 &  &  &  &  &  & 0.12^{***} & 0.10^{**}  & 0.12^{***} & 0.04    & 0.12^{***} &  0.04    &  0.04    &  0.01    & 0.11^{**}  & 0.12^{**}  & 0.03    \\ 
  SNIP & 24 &  &  &  &  &  &  & 0.77^{***} & 0.61^{***} & 0.44^{***} & 0.34    &  0.44^{**}  &  0.37    &  0.28    & 0.35^{*}   & 0.35^{*}   & 0.37    \\ 
  IPP & 24 &  &  &  &  &  &  &  & 0.60^{***} & 0.44^{***} & 0.34    &  0.44^{**}  &  0.36    &  0.28    & 0.35^{*}   & 0.35^{*}   & 0.37    \\ 
  SJR & 24 &  &  &  &  &  &  &  &  & 0.44^{***} & 0.34    &  0.44^{**}  &  0.37    &  0.28    & 0.34^{*}   & 0.35^{*}   & 0.37    \\ 
  Wie01 & 22 &  &  &  &  &  &  &  &  &  & 0.32    &  0.17    &  0.27    &  0.02    & 0.25    & 0.23    & 0.19    \\ 
  Vhb03 & 28 &  &  &  &  &  &  &  &  &  &  &  0.10    &  0.23    &  0.01    & 0.47^{***} & 0.44^{***} & 0.18    \\ 
  Bjm04 & 16 &  &  &  &  &  &  &  &  &  &  &  &  0.13    & -0.03    & 0.06    & 0.06    & 0.16    \\ 
  Hkb05 & 21 &  &  &  &  &  &  &  &  &  &  &  &  &  0.04    & 0.22    & 0.20    & 0.21    \\ 
  Theo05 & 6 &  &  &  &  &  &  &  &  &  &  &  &  &  & 0.02    & 0.02    & 0.11    \\ 
  Ejis07 & 31 &  &  &  &  &  &  &  &  &  &  &  &  &  &  & 0.82^{***} & 0.23    \\ \hline
& $N_R$ & \multicolumn{1}{l}{Ast08} & \multicolumn{1}{l}{Wie08} & \multicolumn{1}{l}{ABS10} & \multicolumn{1}{l}{Den11} & \multicolumn{1}{l}{HEC11} & \multicolumn{1}{l}{UQ11}& \multicolumn{1}{l}{Vhb11} & \multicolumn{1}{l}{Aeres12} & \multicolumn{1}{l}{Cra12} &\multicolumn{1}{l}{EJL12} & \multicolumn{1}{l}{Abdc13} & \multicolumn{1}{l}{Cnrs13} & \multicolumn{1}{l}{Ess13} & \multicolumn{1}{l}{Fneg13}& \multicolumn{1}{l}{MLE}& \\ 

  \hline
IF & 58 & 0.03    & 0.01    & 0.09^{**}  & 0.09^{*}   & 0.03    & 0.02    & 0.04    & 0.03    & 0.04^{*}   & 0.06^{*}   &  0.10^{***} & 0.05    & 0.04    & 0.00    & 0.62^{***} &  \\ 
  IF5Y & 56 & 0.04    & 0.04    & 0.10^{**}  & 0.10^{**}  & 0.05    & 0.03    & 0.06    & 0.03    & 0.06^{**}  & 0.09^{**}  &  0.12^{***} & 0.05    & 0.05    & 0.01    & 0.63^{***} &  \\ 
  IMMI & 56 & 0.02    & 0.00    & 0.06    & 0.07    & 0.05    & 0.00    & 0.03    & 0.03    & 0.02    & 0.10^{**}  &  0.08^{*}   & 0.06    & 0.02    & 0.01    & 0.49^{***} &  \\ 
  CHL & 57 & 0.02    & 0.04    & 0.00    & 0.02    & 0.02    & 0.02    & 0.02    & 0.02    & 0.01    & 0.02    & -0.01    & 0.03    & 0.02    & 0.01    & 0.22^{**}  &  \\ 
  EFS & 58 & 0.04^{*}   & 0.02    & 0.05    & 0.06    & 0.04    & 0.02    & 0.05    & 0.06^{**}  & 0.02    & 0.05    &  0.10^{**}  & 0.09^{***} & 0.03    & 0.01    & 0.68^{***} &  \\ 
  AIS & 56 & 0.04    & 0.05    & 0.10^{**}  & 0.13^{***} & 0.07^{*}   & 0.04    & 0.09^{**}  & 0.03    & 0.07^{**}  & 0.09^{**}  &  0.10^{**}  & 0.08^{*}   & 0.06^{*}   & 0.01    & 0.60^{***} &  \\ 
  SNIP & 24 & 0.38^{*}   & 0.36^{*}   & 0.33    & 0.32^{*}   & 0.38^{*}   & 0.38^{*}   & 0.32    & 0.38^{*}   & 0.40^{*}   & 0.34    &  0.33    & 0.36^{*}   & 0.33    & 0.33    & 0.10^{***} &  \\ 
  IPP & 24 & 0.38^{*}   & 0.36^{*}   & 0.33    & 0.32^{*}   & 0.38^{*}   & 0.38^{*}   & 0.33    & 0.38^{*}   & 0.40^{*}   & 0.34    &  0.33    & 0.36^{*}   & 0.33    & 0.34    & 0.10^{***} &  \\ 
  SJR & 24 & 0.38^{*}   & 0.36^{*}   & 0.33    & 0.32^{*}   & 0.38^{*}   & 0.38^{*}   & 0.33    & 0.38^{*}   & 0.40^{*}   & 0.35    &  0.32    & 0.36^{*}   & 0.33    & 0.34    & 0.11^{***} &  \\ 
  Wie01 & 22 & 0.18    & 0.30    & 0.16    & 0.19    & 0.25    & 0.18    & 0.17    & 0.30    & 0.16    & 0.23    &  0.18    & 0.24    & 0.21    & 0.23    & 0.05^{**}  &  \\ 
  Vhb03 & 28 & 0.11    & 0.36^{**}  & 0.27    & 0.22    & 0.23    & 0.17    & 0.40^{***} & 0.25    & 0.19    & 0.22    &  0.19    & 0.29    & 0.25    & 0.18    & 0.09^{***} &  \\ 
  Bjm04 & 16 & 0.23    & 0.13    & 0.07    & 0.04    & 0.18    & 0.19    & 0.09    & 0.10    & 0.10    & 0.15    &  0.02    & 0.16    & 0.16    & 0.28    & 0.01    &  \\ 
  Hkb05 & 21 & 0.14    & 0.25    & 0.14    & 0.08    & 0.23    & 0.21    & 0.20    & 0.22    & 0.22    & 0.17    &  0.06    & 0.19    & 0.17    & 0.28    & 0.04^{*}   &  \\ 
  Theo05 & 6 & 0.03    & 0.02    & 0.02    & 0.01    & 0.01    & 0.11    & 0.07    & 0.12    & 0.06    & 0.02    &  0.01    & 0.07    & 0.07    & 0.35    & 0.01    &  \\ 
  Ejis07 & 31 & 0.10    & 0.25    & 0.21    & 0.30^{*}   & 0.22    & 0.14    & 0.33^{**}  & 0.22    & 0.15    & 0.21    &  0.23    & 0.30    & 0.17    & 0.08    & 0.14^{***} &  \\ 
  EjisCI07 & 32 & 0.12    & 0.30    & 0.23    & 0.37^{**}  & 0.21    & 0.13    & 0.34^{**}  & 0.22    & 0.16    & 0.23    &  0.25    & 0.29    & 0.22    & 0.08    & 0.14^{***} &  \\ 
  UQ07 & 21 & 0.25    & 0.22    & 0.18    & 0.10    & 0.20    & 0.47^{***} & 0.23    & 0.22    & 0.25    & 0.19    &  0.23    & 0.25    & 0.17    & 0.09    & 0.05^{**}  &  \\ 
  Ast08 & 19 &  & 0.20    & 0.19    & 0.09    & 0.22    & 0.27    & 0.16    & 0.22    & 0.30    & 0.19    &  0.23    & 0.22    & 0.19    & 0.18    & 0.05^{**}  &  \\ 
  Wie08 & 25 &  &  & 0.28    & 0.26    & 0.21    & 0.22    & 0.28    & 0.30    & 0.24    & 0.24    &  0.20    & 0.24    & 0.36    & 0.27    & 0.03^{*}   &  \\ 
  ABS10 & 28 &  &  &  & 0.32^{*}   & 0.20    & 0.18    & 0.22    & 0.21    & 0.25    & 0.28    &  0.29    & 0.23    & 0.48^{***} & 0.11    & 0.09^{***} &  \\ 
  Den11 & 34 &  &  &  &  & 0.12    & 0.13    & 0.21    & 0.21    & 0.12    & 0.25    &  0.34^{**}  & 0.20    & 0.24    & 0.06    & 0.12^{**}  &  \\ 
  HEC11 & 22 &  &  &  &  &  & 0.15    & 0.23    & 0.18    & 0.26    & 0.30    &  0.20    & 0.27    & 0.29    & 0.22    & 0.06^{**}  &  \\ 
  UQ11 & 19 &  &  &  &  &  &  & 0.17    & 0.22    & 0.25    & 0.14    &  0.20    & 0.23    & 0.17    & 0.12    & 0.03^{*}   &  \\ 
  Vhb11 & 26 &  &  &  &  &  &  &  & 0.29    & 0.23    & 0.26    &  0.17    & 0.32    & 0.24    & 0.21    & 0.09^{***} &  \\ 
  Aeres12 & 24 &  &  &  &  &  &  &  &  & 0.22    & 0.21    &  0.26    & 0.65^{***} & 0.25    & 0.15    & 0.05^{**}  &  \\ 
  Cra12 & 18 &  &  &  &  &  &  &  &  &  & 0.22    &  0.22    & 0.23    & 0.25    & 0.20    & 0.06^{***} &  \\ 
  EJL12 & 26 &  &  &  &  &  &  &  &  &  &  &  0.24    & 0.34    & 0.27    & 0.15    & 0.10^{***} &  \\ 
  Abdc13 & 29 &  &  &  &  &  &  &  &  &  &  &  & 0.27    & 0.19    & 0.06    & 0.12^{***} &  \\ 
  Cnrs13 & 26 &  &  &  &  &  &  &  &  &  &  &  &  & 0.29    & 0.15    & 0.11^{***} &  \\ 
  Ess13 & 25 &  &  &  &  &  &  &  &  &  &  &  &  &  & 0.19    & 0.05^{*}   &  \\ 
  Fneg13 & 11 &  &  &  &  &  &  &  &  &  &  &  &  &  &  & 0.02^{**}  &  \\ \hline
\label{tab:rstats}
\end{tabular} 
\begin{minipage}{\linewidth}
  \vspace{-0.5cm}
 \renewcommand{\footnoterule}{}
 \footnotetext{$^{***}$ -- $p$ value $< 0.001$, $^{**}$ -- $p$ value $< 0.01$, $^{*}$-- $p$ value $< 0.05$. }
\end{minipage}
\end{tiny}
\end{sidewaystable} 

\section{Methodological approach}\label{sect:model}
Consider a total number of $K$ journal rankings $\mathcal{R} = \{R_1, \dots, R_K\}$ of $N$
journals $\mathcal{J} = \{J_1 ,\dots, J_N \}$. Each journal in the set $\mathcal{J}$ is ranked in at least one of the rankings in $\mathcal{R}$.

The aim of the analysis is suitably aggregating the information available from the $K$ different ranking sources by finding a metric that reflects the overall quality of the journals in $\mathcal{J}$.
For each ranking $R_{k}$, we compare the journals included in the ranking pairwise by
investigating whether a journal $J_i$ has been ranked higher/lower than or the same as another journal $J_j$.
Let $Y_{ijk}$ be a variable taking one of the following values:
\begin{displaymath}
Y_{ijk} =  \begin{cases}
1 & \mbox{if ranking $R_{k}$ rates journal $J_{i}$ higher than journal $J_{j}$},\\
0.5 & \mbox{if ranking $R_{k}$ rates journal $J_{i}$ and journal $J_{j}$ the same},\\
0 & \mbox{if ranking $R_{k}$ rates journal $J_{i}$ lower than journal $J_{j}$}.\\
\end{cases}
\end{displaymath}
A popular statistical model for modeling paired comparison data is the Bradley-Terry model \citep{bradley1952rank}, which is a logistic regression for a binary outcome, where the probability of journal $J_i$ being better than journal $J_j$ is given by $\mathrm{exp}(\mu_i - \mu_j)/(1 + \mathrm{exp}(\mu_i - \mu_j))$ with $\mu_i$ and $\mu_j$ being the ability (quality) parameters for journal $J_i$ and journal $J_j$. 
Note that $N$ need not be higher than $K$.

As many journals in the analysis fall into the same ranking class, we employ a modification of the original Bradley-Terry model which can properly handle ties and treat ties as one-half of a success and one-half of a failure. The log-likelihood is then given by:
\begin{equation}\label{btm}
\ell(\bm{\mu})= \sum_{i<j}^N\sum_{k}^K y_{ijk} (\mu_i - \mu_j) - \text{log}(1 + \mathrm{exp}(\mu_i - \mu_j)).
\end{equation}
The parameter of interest is the vector of journal abilities $\bm{\mu} = (\mu_1, \dots, \mu_N)^\top$ which can be further used for ranking the $N$ journals. Only pairwise differences $\mu_i - \mu_j$ are identifiable in model \eqref{btm}, hence a restriction needs to be imposed on the vector $\bm{\mu}$. We choose to set the ability parameter  $\mu_1$ of the first journal \emph{4OR} equal to zero. 
An alternative to treating ties as one-half of a success and one-half of a failure would be using the cumulative link Bradley-Terry model \citep{agresti2010analysis}. We choose the former due to reduced computational expenses.

In order to avoid potential over interpretation of insignificant differences between journal abilities and obtain a clustering of the journals whose ability parameters are similar, \cite{masarotto2012ranking} propose the ranking lasso technique, which computes the solution for the modified Bradley-Terry model by maximizing the likelihood and imposing a $\mathcal{L}_1$ penalty on the pairwise ability differences $\mu_i-\mu_j$:
\begin{equation}\label{rlasso}
\widehat{\bm{\mu}}_\lambda = \text{arg min} \left\{- \ell(\bm{\mu}) + \lambda\sum_{i<j}^{N} w_{ij} |\mu_i-\mu_j| \right\},
\end{equation}
where $\ell(\bm{\mu})$ is the log-likelihood function in \eqref{btm}, $\lambda$ is the lasso penalty and  $w_{ij}$ are pair specific weights.
We follow the suggestion of \cite{masarotto2012ranking} and choose the weights inversely proportional to the maximum likelihood estimates (MLE): 
\begin{displaymath}
w_{ij} = |\hat{\mu}_i^{\text{(MLE)}} - \hat{\mu}_j^{\text{(MLE)}}|^{-1}.
\end{displaymath}
This method is called the adaptive ranking lasso and it overcomes the problem of inconsistent non-zero lasso estimates due to the severe  $\mathcal{L}_1$ penalty \citep{zou2006adaptive}.
The lasso penalty $\lambda$ can be chosen by minimization of an information criterion such as the Akaike Information Criterion (AIC).

The uncertainty of the estimated abilities is evaluated using parametric bootstrap samples. Resampling from the modified Bradley-Terry model was performed by replacing the true parameters by their estimates. As adaptive lasso estimators (ALASSO) are by construction biased, we use the bias-corrected percentile method \citep{efron1987better} for computing $100(1-2\alpha)\%$ confidence intervals for the ability parameters.

\section{Results}\label{sect:results}       
We proceed with presenting results from applying the method in Section~\ref{sect:model} to the $58$ journals and $31$ rankings in Section \ref{sec:results-all}. In Section~\ref{sec:results2013} we repeat the analysis but use only rankings from 2013 and check the stability of the results. 

\subsection{All rankings}
\label{sec:results-all}

We compute the percentage of times each journal appeared in the rankings in Table~\ref{tab:ranking}. The journal that is rated most often is \emph{Operations Research (OR)}, followed by \emph{Production and Operations Management (POM)} and \emph{Decision Support Systems (DSS)} (see Table~\ref{tab:results}).  Table~\ref{tab:results} also presents the MLEs and the ALASSO estimators where the shrinkage parameter $\lambda$ was chosen according to AIC. The ALASSO estimators group journals with similar ability parameters and indicate 24 clusters of journals. 
In addition, 95\% bias-corrected confidence intervals are reported. The uncertainty in the ability parameters was estimated using 1000 parametric bootstrap samples. An important feature that can be observed from the confidence intervals for the MLEs is that they are overlapping in some cases. This implies that some of the differences between the ability parameters are not significant and that the ranking lasso method should be employed for shrinking the differences of similar coefficients to zero. In most of the cases, the confidence intervals for the ALASSO estimators are shorter than for MLE, indicating that the shrinkage provides a reduction in variability.

The correlation between the MLE and rankings used in the analysis is positive (see column ``MLE'' in Table~\ref{tab:rstats}).
 The metric exhibiting the highest correlation with the MLE ranking of the modified Bradley-Terry model is the \emph{Eigenfactor\textsuperscript{\textregistered} Score}.
\begin{landscape}
\begin{scriptsize}  

\setlength{\LTcapwidth}{8in}
\begin{longtable}{@{\extracolsep{\fill}}llcrcrc|ccrcrc}
\caption[.]{Maximum likelihood (MLE) and adaptive lasso estimators (ALASSO) for the modified Bradley-Terry model, with corresponding 95\% bias-corrected percentile bootstrap confidence intervals}\\
\hline
 \multicolumn{7}{c|}{Rankings 2001--2013} &   \multicolumn{6}{c}{Rankings 2013}\\   
 Pos&  Journal  &\% rated& MLE&95\% c.i. & ALASSO & 95\% c.i. & Pos& \% rated& MLE&95\% c.i.  & ALASSO& 95\% c.i.\\
 \hline
 \endfirsthead
     \caption[]{(continued)}\\\hline
    \multicolumn{7}{c|}{Rankings 2001--2013} &   \multicolumn{6}{c}{Rankings 2013}\\   
 Pos&  Journal  &\% rated& MLE&95\% c.i. & ALASSO & 95\% c.i. & Pos& \% rated& MLE&95\% c.i.  & ALASSO& 95\% c.i.\\
 \hline
\endhead
\hline
\endfoot
    1 & MS & 83.9 & 3.500 & (3.239, 3.753) & 2.980 & (2.900, 3.252) &    1 & 76.9 & 3.790 & (3.486, 4.144) & 2.730 & (2.648, 3.159) \\ 
     2 & OR & 96.8 & 2.420 & (2.193, 2.634) & 1.930 & (1.858, 2.176) &    8 & 92.3 & 2.050 & (1.791, 2.288) & 1.400 & (1.339, 1.621) \\ 
     3 & MP & 64.5 & 2.390 & (2.165, 2.604) & 1.930 & (1.846, 2.176) &    4 & 61.5 & 2.530 & (2.268, 2.790) & 1.680 & (1.586, 1.904) \\ 
     4 & TRBM & 58.1 & 2.340 & (2.119, 2.557) & 1.930 & (1.848, 2.176) &    3 & 61.5 & 2.780 & (2.470, 3.035) & 1.880 & (1.835, 2.209) \\ 
     5 & JOM & 87.1 & 2.300 & (2.093, 2.503) & 1.930 & (1.863, 2.176) &    2 & 76.9 & 2.850 & (2.586, 3.116) & 1.880 & (1.853, 2.199) \\ 
     6 & TS & 61.3 & 2.060 & (1.848, 2.298) & 1.680 & (1.568, 2.022) &    5 & 69.2 & 2.250 & (1.992, 2.491) & 1.400 & (1.339, 1.606) \\ 
     7 & SCL & 35.5 & 1.790 & (1.557, 2.023) & 1.310 & (1.225, 1.561) &    6 & 46.2 & 2.220 & (1.930, 2.466) & 1.400 & (1.349, 1.606) \\ 
     8 & EJOR & 80.6 & 1.650 & (1.438, 1.842) & 1.270 & (1.186, 1.491) &    7 & 69.2 & 2.050 & (1.831, 2.334) & 1.400 & (1.345, 1.796) \\ 
     9 & TRELT & 64.5 & 1.480 & (1.268, 1.687) & 1.040 & (0.973, 1.481) &   11 & 92.3 & 1.690 & (1.449, 1.902) & 1.200 & (1.061, 1.499) \\ 
    10 & POM & 93.5 & 1.400 & (1.205, 1.605) & 1.040 & (0.963, 1.351) &   10 & 100.0 & 1.750 & (1.518, 1.960) & 1.200 & (1.063, 1.507) \\ 
    11 & MOR & 77.4 & 1.320 & (1.120, 1.528) & 0.970 & (0.888, 1.187) &   16 & 92.3 & 1.210 & (0.994, 1.437) & 0.770 & (0.691, 1.025) \\ 
    12 & DSS & 90.3 & 1.300 & (1.098, 1.493) & 0.970 & (0.890, 1.202) &   14 & 76.9 & 1.310 & (1.095, 1.570) & 0.770 & (0.692, 1.014) \\ 
    13 & MSOM & 64.5 & 1.170 & (0.981, 1.394) & 0.870 & (0.776, 1.158) &   13 & 100.0 & 1.360 & (1.147, 1.580) & 0.770 & (0.693, 1.079) \\ 
    14 & JOTA & 51.6 & 1.160 & (0.959, 1.390) & 0.870 & (0.775, 1.158) &   12 & 53.8 & 1.450 & (1.194, 1.676) & 0.770 & (0.690, 1.030) \\ 
    15 & OMEGA & 87.1 & 1.150 & (0.947, 1.356) & 0.870 & (0.766, 1.186) &    9 & 69.2 & 1.980 & (1.742, 2.224) & 1.400 & (1.340, 1.604) \\ 
    16 & JGO & 19.4 & 0.920 & (0.686, 1.157) & 0.560 & (0.494, 0.944) &   18 & 46.2 & 0.970 & (0.732, 1.193) & 0.420 & (0.310, 0.760) \\ 
    17 & IIE & 80.6 & 0.910 & (0.720, 1.108) & 0.560 & (0.501, 0.755) &   23 & 84.6 & 0.620 & (0.395, 0.831) & 0.370 & (0.234, 0.662) \\ 
    18 & IJPR & 87.1 & 0.870 & (0.685, 1.082) & 0.560 & (0.507, 0.812) &   19 & 76.9 & 0.890 & (0.668, 1.113) & 0.380 & (0.243, 0.653) \\ 
    19 & JSCH & 61.3 & 0.860 & (0.631, 1.054) & 0.560 & (0.504, 0.755) &   28 & 84.6 & 0.480 & (0.261, 0.698) & 0.370 & (0.241, 0.625) \\ 
    20 & INFORMS & 61.3 & 0.840 & (0.637, 1.053) & 0.560 & (0.499, 0.766) &   22 & 84.6 & 0.750 & (0.534, 0.956) & 0.380 & (0.230, 0.628) \\ 
    21 & EXSA & 54.8 & 0.740 & (0.535, 0.961) & 0.490 & (0.396, 0.754) &   20 & 69.2 & 0.810 & (0.580, 1.028) & 0.380 & (0.244, 0.626) \\ 
    22 & AOR & 83.9 & 0.710 & (0.503, 0.932) & 0.490 & (0.395, 0.769) &   21 & 69.2 & 0.760 & (0.521, 0.974) & 0.380 & (0.242, 0.626) \\ 
    23 & COR & 77.4 & 0.690 & (0.492, 0.891) & 0.490 & (0.398, 0.754) &   15 & 69.2 & 1.250 & (1.002, 1.472) & 0.770 & (0.697, 1.134) \\ 
    24 & ORS & 64.5 & 0.600 & (0.397, 0.802) & 0.450 & (0.302, 0.692) &   26 & 69.2 & 0.530 & (0.321, 0.760) & 0.370 & (0.241, 0.625) \\ 
    25 & JORS & 83.9 & 0.590 & (0.390, 0.806) & 0.450 & (0.305, 0.696) &   25 & 69.2 & 0.560 & (0.319, 0.803) & 0.370 & (0.239, 0.653) \\ 
    26 & COA & 19.4 & 0.570 & (0.337, 0.816) & 0.450 & (0.304, 0.708) &   24 & 46.2 & 0.610 & (0.348, 0.830) & 0.370 & (0.243, 0.653) \\ 
    27 & ORL & 74.2 & 0.510 & (0.318, 0.718) & 0.450 & (0.304, 0.683) &   32 & 69.2 & 0.110 & ($-$0.121, 0.334) & 0.000 & ($-$0.430, 0.000) \\ 
    28 & NSE & 29.0 & 0.490 & (0.250, 0.723) & 0.450 & (0.297, 0.683) &   27 & 69.2 & 0.510 & (0.303, 0.743) & 0.370 & (0.242, 0.653) \\ 
    29 & TNV & 74.2 & 0.410 & (0.206, 0.601) & 0.380 & (0.072, 0.526) &   17 & 69.2 & 1.100 & (0.882, 1.320) & 0.720 & (0.607, 0.935) \\ 
    30 & NETW & 38.7 & 0.360 & (0.142, 0.587) & 0.380 & (0.071, 0.527) &   31 & 53.8 & 0.240 & (0.020, 0.490) & 0.270 & (0.000, 0.488) \\ 
    31 & OMS & 29.0 & 0.300 & (0.045, 0.526) & 0.380 & (0.063, 0.525) &   30 & 69.2 & 0.310 & (0.080, 0.538) & 0.270 & (0.000, 0.498) \\ 
    32 & NAVRL & 71.0 & 0.110 & ($-$0.086, 0.327) & 0.000 & ($-$0.459, 0.000) &   36 & 92.3 & $-$0.200 & ($-$0.424, 0.012) & $-$0.280 & ($-$0.540, $-$0.121) \\ 
    33 & JMS & 54.8 & 0.040 & ($-$0.165, 0.254) & 0.000 & ($-$0.424, 0.000) &   29 & 69.2 & 0.330 & (0.104, 0.555) & 0.270 & (0.000, 0.490) \\ 
    34 & FODM & 29.0 & 0.010 & ($-$0.214, 0.234) & 0.000 & ($-$0.459, 0.000) &   33 & 69.2 & 0.010 & ($-$0.209, 0.228) & 0.000 & ($-$0.436, 0.000) \\ 
    35 & 4OR & 29.0 & 0.000 &  & 0.000 &  &   34 & 69.2 & 0.000 &  & 0.000 &  \\ 
    36 & EO & 29.0 & $-$0.060 & ($-$0.281, 0.190) & 0.000 & ($-$0.455, 0.000) &   35 & 69.2 & $-$0.060 & ($-$0.303, 0.148) & 0.000 & ($-$0.445, 0.000) \\ 
    37 & PPC & 80.6 & $-$0.230 & ($-$0.419, $-$0.023) & $-$0.300 & ($-$0.624, $-$0.182) &   39 & 76.9 & $-$0.380 & ($-$0.601, $-$0.134) & $-$0.280 & ($-$0.544, $-$0.113) \\ 
    38 & QS & 35.5 & $-$0.280 & ($-$0.492, $-$0.040) & $-$0.300 & ($-$0.629, $-$0.184) &   38 & 46.2 & $-$0.290 & ($-$0.535, $-$0.060) & $-$0.280 & ($-$0.538, $-$0.118) \\ 
    39 & OPT & 19.4 & $-$0.280 & ($-$0.520, $-$0.038) & $-$0.300 & ($-$0.650, $-$0.183) &   37 & 46.2 & $-$0.280 & ($-$0.496, $-$0.054) & $-$0.280 & ($-$0.540, $-$0.118) \\ 
    40 & INTER & 77.4 & $-$0.310 & ($-$0.517, $-$0.091) & $-$0.300 & ($-$0.561, $-$0.184) &   43 & 53.8 & $-$0.630 & ($-$0.906, $-$0.396) & $-$0.330 & ($-$0.546, $-$0.212) \\ 
    41 & OCAM & 19.4 & $-$0.420 & ($-$0.647, $-$0.176) & $-$0.300 & ($-$0.633, $-$0.180) &   40 & 46.2 & $-$0.420 & ($-$0.676, $-$0.206) & $-$0.280 & ($-$0.544, $-$0.117) \\ 
    42 & OL & 29.0 & $-$0.480 & ($-$0.706, $-$0.265) & $-$0.330 & ($-$0.655, $-$0.245) &   41 & 69.2 & $-$0.500 & ($-$0.734, $-$0.271) & $-$0.300 & ($-$0.561, $-$0.151) \\ 
    43 & ITOR & 64.5 & $-$0.550 & ($-$0.760, $-$0.323) & $-$0.370 & ($-$0.658, $-$0.286) &   47 & 53.8 & $-$0.840 & ($-$1.096, $-$0.582) & $-$0.330 & ($-$0.536, $-$0.210) \\ 
    44 & DEDS & 19.4 & $-$0.620 & ($-$0.878, $-$0.373) & $-$0.370 & ($-$0.658, $-$0.295) &   42 & 46.2 & $-$0.620 & ($-$0.861, $-$0.380) & $-$0.330 & ($-$0.562, $-$0.215) \\ 
    45 & DO & 29.0 & $-$0.620 & ($-$0.853, $-$0.401) & $-$0.370 & ($-$0.657, $-$0.288) &   44 & 69.2 & $-$0.640 & ($-$0.884, $-$0.424) & $-$0.330 & ($-$0.562, $-$0.224) \\ 
    46 & IJITD & 19.4 & $-$0.700 & ($-$0.934, $-$0.438) & $-$0.370 & ($-$0.658, $-$0.301) &   45 & 46.2 & $-$0.710 & ($-$0.961, $-$0.468) & $-$0.330 & ($-$0.562, $-$0.214) \\ 
    47 & OE & 29.0 & $-$0.780 & ($-$1.002, $-$0.542) & $-$0.370 & ($-$0.658, $-$0.293) &   46 & 69.2 & $-$0.800 & ($-$1.020, $-$0.561) & $-$0.330 & ($-$0.546, $-$0.226) \\ 
    48 & MMOR & 67.7 & $-$0.920 & ($-$1.126, $-$0.706) & $-$0.740 & ($-$1.102, $-$0.604) &   49 & 84.6 & $-$1.230 & ($-$1.472, $-$0.986) & $-$0.850 & ($-$1.173, $-$0.736) \\ 
    49 & TOP & 29.0 & $-$1.190 & ($-$1.410, $-$0.932) & $-$0.890 & ($-$1.172, $-$0.813) &   48 & 69.2 & $-$1.220 & ($-$1.481, $-$0.985) & $-$0.850 & ($-$1.173, $-$0.723) \\ 
    50 & PJO & 29.0 & $-$1.230 & ($-$1.481, $-$0.980) & $-$0.890 & ($-$1.172, $-$0.809) &   50 & 69.2 & $-$1.260 & ($-$1.511, $-$1.032) & $-$0.850 & ($-$1.053, $-$0.736) \\ 
    51 & CEJOR & 29.0 & $-$1.250 & ($-$1.501, $-$1.001) & $-$0.890 & ($-$1.138, $-$0.826) &   51 & 69.2 & $-$1.290 & ($-$1.554, $-$1.055) & $-$0.850 & ($-$1.141, $-$0.731) \\ 
    52 & INFOR & 16.1 & $-$1.410 & ($-$1.698, $-$1.114) & $-$0.890 & ($-$1.222, $-$0.798) &   52 & 38.5 & $-$1.430 & ($-$1.719, $-$1.152) & $-$0.850 & ($-$1.090, $-$0.724) \\ 
    53 & JSIM & 22.6 & $-$1.650 & ($-$1.905, $-$1.359) & $-$1.230 & ($-$1.498, $-$1.142) &   53 & 53.8 & $-$1.690 & ($-$1.970, $-$1.410) & $-$1.140 & ($-$1.339, $-$1.090) \\ 
    54 & ASMB & 29.0 & $-$1.790 & ($-$2.040, $-$1.521) & $-$1.260 & ($-$1.530, $-$1.199) &   55 & 69.2 & $-$1.830 & ($-$2.077, $-$1.561) & $-$1.140 & ($-$1.364, $-$1.088) \\ 
    55 & RAIRO & 19.4 & $-$1.800 & ($-$2.082, $-$1.517) & $-$1.260 & ($-$1.495, $-$1.217) &   54 & 46.2 & $-$1.830 & ($-$2.108, $-$1.556) & $-$1.140 & ($-$1.401, $-$1.072) \\ 
    56 & IMAJMM & 29.0 & $-$2.000 & ($-$2.264, $-$1.736) & $-$1.390 & ($-$1.740, $-$1.346) &   56 & 69.2 & $-$2.040 & ($-$2.329, $-$1.781) & $-$1.270 & ($-$1.472, $-$1.222) \\ 
    57 & APJOR & 19.4 & $-$2.860 & ($-$3.207, $-$2.489) & $-$2.120 & ($-$2.430, $-$2.038) &   57 & 46.2 & $-$2.890 & ($-$3.285, $-$2.526) & $-$1.980 & ($-$2.300, $-$1.900) \\ 
    58 & MILOR & 12.9 & $-$5.160 & ($-$6.653, $-$4.329) & $-$3.320 & ($-$3.788, $-$3.243) &   58 & 30.8 & $-$5.200 & ($-$6.649, $-$4.327) & $-$3.050 & ($-$3.545, $-$2.986) \\ \label{tab:results}
\end{longtable} 
\end{scriptsize}
\end{landscape} 

\begin{figure}
  \caption{Journal quality scores based on the whole sample of rankings -- ALASSO estimators}
\vspace{-1cm}
\includegraphics{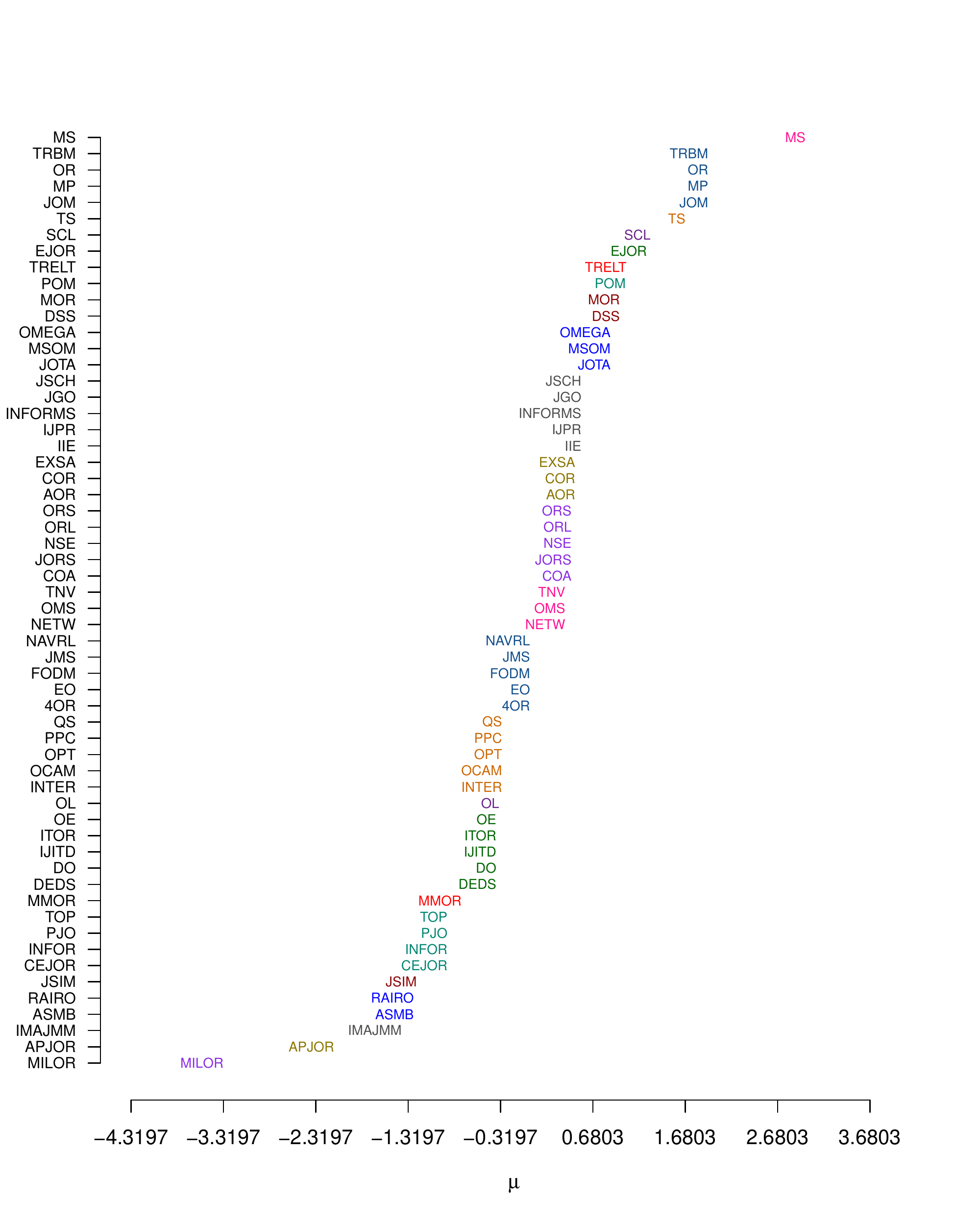}
\label{fig:aic}
\end{figure}

Figure~\ref{fig:aic} illustrates the ALASSO estimators for each journal. Journals clustered together are marked by the same coloring. The journal with the highest ability parameter is \emph{Management Science} followed by \emph{Operations Research}, \emph{Mathematical Programming (MP)}, \emph{Transportation Research Part B: Methodological} and \emph{Journal of Operations Management (JOM)}. The ranking lasso groups the latter four journals into one cluster because of their similar MLEs. The journal with the lowest score is \emph{Military Operations Research (MILOR)}.

\cite{petersen2011journal}, \cite{fry2013outlets}, and \cite{tuselmann2015towards} also identify \emph{Management Science} as the top outlet for OR/MS/POM research. When regarding \emph{only} the intersection of the sets of journals, \emph{JOM} and \emph{POM} (position 5 and 10 in the MLE ranking) also make it in the top ten lists of  \cite{meredith2011knowledge} (position 1 and 2 respectively), \cite{petersen2011journal} (position 3 and 4), \cite{fry2013outlets} (position 3 and 4). In \cite{tuselmann2015towards}, \emph{JOM} is ranked second in the OR/MS/POM subject area while \emph{POM} is on position 9. \emph{OR, MP, TS, EJOR} are in the top ten of \cite{fry2013outlets}, \cite{tuselmann2015towards}. \emph{IJPR} receives a top ten position in \cite{meredith2011knowledge}, \cite{petersen2011journal} and \cite{fry2013outlets}, but in our analysis it is ranked only on position 18 \citep[and position 11 in][]{tuselmann2015towards}.

\setkeys{Gin}{width=\textwidth}

\subsection{Using 2013 rankings only}
\label{sec:results2013}

\begin{figure}
  \caption{Journal quality scores based on the rankings dated in 2013 -- ALASSO estimators}
\vspace{-1cm}
\includegraphics{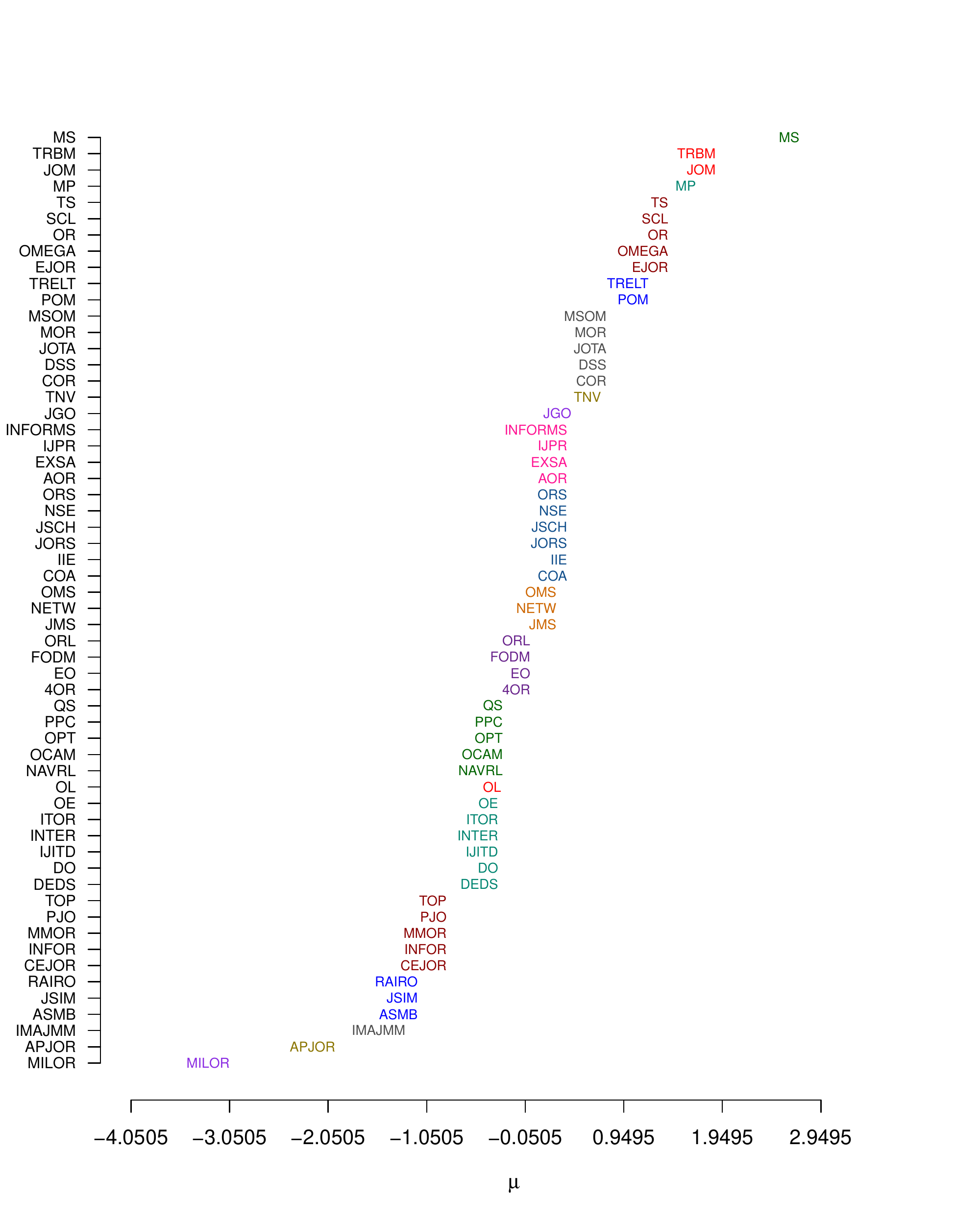}
\label{fig:aic2013}
\end{figure}

An aspect that can be verified in our proposed framework is whether the new rankings (i.e., dated 2013) lead to a different consensus ranking from the one obtained using the whole sample of rankings. Rankings complied in 2013 could provide a more up-to-date picture of the current research environment. We address this topic be redoing the analysis using only the 13 rankings published in 2013. The MLE and ALASSO estimates together with 95\% confidence intervals are presented in Table~\ref{tab:results}. ALASSO estimators are visualized in Figure~\ref{fig:aic2013}.
\setkeys{Gin}{width=\textwidth}
The resulting journal ability parameters are similar to the ones using the whole sample of 31 rankings and the consensus ranking remains highly stable. However, few differences can be identified. On the one hand, \emph{OR} is downgraded from the second cluster to the fourth cluster and the estimated ability parameters have decreased significantly. Significantly reduced ALASSO estimators are observed also for \emph{ORL} and \emph{NAVRL}. On the other hand, \emph{JOM}, \emph{OMEGA}, \emph{TNV} and \emph{COR} have higher estimated ability parameters and higher positions in the ranking.

\section{Conclusion}\label{sect:concl}

We use the ranking lasso method proposed by \cite{masarotto2012ranking} together with the Bradley-Terry model for paired comparison data in order to estimate quality scores for 58 OR/MS journals. The approach overcomes the issue of missing data by using pairwise comparisons and allows the identification of clusters of journals with similar quality scores. We conclude that \emph{Management Science} maintains its position as lead OR/MS outlet, followed by \emph{OR, MP, TRBM, JOM, TS, SCL, EJOR, TRELT} and \emph{POM}. The meta-analysis using only 2013 rankings indicates increased quality for \emph{OMEGA} and \emph{TNV}. 
The presented method is flexible to integrate a wide range of heterogeneous rankings and the resulting graph provides a detailed outline on how to cluster and/or rate the underlying journals for subsequent usage for different purposes.



\begin{appendices}

\section{Journal abbreviations and ranking lists}

\begin{sidewaystable}
  \begin{scriptsize}  
  \caption[.]{Journal abbreviations.}
    \begin{tabularx}{\linewidth}{p{1.2cm}Xp{1cm}X}
      \hline
    Abbrev & Journal&Abbrev & Journal \\ 
    \hline
  4OR & 4OR -- A Quarterly Journal of Operations Research &  MILOR &  Military Operations Research\\ 
  AOR & Annals of Operations Research& MMOR & Mathematical Methods of Operations Research \\ 
  APJOR & Asia Pacific Journal of Operational Research & MOR & Mathematics of Operations Research\\ 
  ASMB & Applied Stochastic Models in Business and Industry&MP & Mathematical Programming \\ 
  CEJOR & Central European Journal of Operations Research& MS & Management Science\\ 
  COA & Computational Optimization and Applications & MSOM & Manufacturing and Service Operations Management \\ 
  COR & Computers \& Operations Research & NAVRL & Naval Research Logistics \\ 
  DEDS &  Discrete Event Dynamic Systems  &  NETW & Networks\\ 
  DO & Discrete Optimization&  NSE & Networks and Spatial Economics \\ 
  DSS & Decision Support Systems & OCAM & Optimal Control Applications and Methods\\ 
  EJOR & European Journal of Operational Research   & OE & Optimization and Engineering\\ 
  EO & Engineering Optimization &   OL & Optimization Letters  \\ 
  EXSA & Expert System with Applications& OMEGA& OMEGA - International Journal of Management Science \\ 
  FODM & Fuzzy Optimization and Decision Making&   OMS & Optimization Methods and Software \\ 
  IIE & IIE Transactions (Institute of Industrial Engineers)&  OPT &  Optimization   \\ 
  IJITD & International Journal of Information Technology \& Decision Making&  OR & Operations Research\\ 
  IJPR & International Journal of Production Research&ORL & Operations Research Letters\\ 
  IMAJMM & IMA Journal Management Mathematics&   ORS & OR Spectrum \\ 
  
  INFOR &  INFOR: Information Systems and Operational Research & PJO & Pacific Journal of Optimization\\ 
 INFORMS & INFORMS Journal on Computing &  POM & Production and Operations Management\\ 
  INTER & Interfaces &PPC & Production Planning \& Control \\ 
  ITOR & International Transactions in Operational Research&   QS & Queueing Systems \\ 
  JGO &Journal of Global Optimization &   RAIRO & RAIRO -- Operations Research \\
  JMS & Journal of Manufacturing Systems&  SCL & Systems \& Control Letters\\ 
  JOM & Journal of Operations Management&  TNV & Technovation \\ 
 JORS & Journal of the Operational Research Society  &  TOP & TOP -- An Official Journal of the Spanish Society of Statistics and Operations Research\\ 
JOTA & Journal of Optimization Theory \& Applications  & TRBM & Transportation Research Part B: Methodological \\ 
 JSCH& Journal of Scheduling  &  TRELT & Transportation Research Part E: Logistics and Transportation Review \\ 
JSIM & Journal of Simulation &   TS & Transportation Science \\ 
\hline
\end{tabularx}
\label{tab:journals}
\end{scriptsize}
\end{sidewaystable} 

\begin{landscape}
  \begin{scriptsize}
\begin{longtable}{@{\extracolsep{\fill}}p{0.05\linewidth}p{0.25\linewidth}p{0.43\linewidth}p{0.05\linewidth}p{0.22\linewidth}}
  \caption[.]{Ranking lists.}\\
  \hline
   Abbrev& Ranking & Source& Year & Scale\\
  \hline
\endfirsthead
 \caption[]{(continued)}\\\hline
  Abbrev& Ranking & Source& Year & Scale\\
  \hline
\endhead
\hline
\endfoot
  IF\footnotemark[1]& Impact Factor &Thomson Reuters& 2013&Numeric from 0. \\ 
  IF5Y\footnotemark[1] & 5-Year Impact Factor &Thomson Reuters& 2013& Numeric from 0\\
  IMMI\footnotemark[1] & Immediacy Index &Thomson Reuters& 2013& Numeric from 0\\ 
  CHL\footnotemark[1] & Cited Half-Life &Thomson Reuters& 2013 &  Numeric from 0\\
   EFS\footnotemark[1]& Eigenfactor\textsuperscript{\textregistered} Score& University of Washington&2013&Numeric from 0 to 1\\
   AIS\footnotemark[1]& Article Influence\textsuperscript{\textregistered} Score& University of Washington& 2013&  Numeric from 0\\
     SNIP\footnotemark[2]& Source Normalized Impact per Paper&Elsevier&2013&  Numeric from 0\\
     IPP\footnotemark[2] & Impact per Publication&Elsevier&2013&  Numeric from 0\\
     SJR\footnotemark[2] &SCImago Journal Rank &Elsevier&2013&  Numeric from 0\\ 
   Wie01 & WU Wien Journal Rating&WU Vienna University of Economics and Business&2001&D < C < B < A < A+\\
  Vhb03 & VHB ranking&Association of University Professors of Business in German speaking countries& 2003 &E < D < C < B < A < A+\\
   Bjm04 & British RAE Rankings&\cite{geary2004journal} & 2004 &Numeric from 1 to 7\\
   Hkb05 & HKBU ranking &Hong Kong Baptist University School of Business& 2005&B- < B < B+ < A\\
   Theo05 &Theoharakis integrated ranking &Theoharakis et al. index&2005&Numeric from 1.2 to 95 \\
   Ejis07& European Journal of Information Systems ranking &\cite{mingers2007ranking}&2007& 1 < 2 < 3 < 4\\
  EjisCI07& European Journal of Information Systems ranking including citation impact factors&\cite{mingers2007ranking}&2007& 1 < 2 < 3 < 4\\    
UQ07 &University of Queensland Rating&University of Queensland& 2007& 5 < 4 < 3 < 2 < 1\\
     Ast08 & Aston ranking&Aston Business School& 2008&1 < 2 < 3 < 4\\
      Wie08 &WU Wien Journal Rating& WU Vienna University of Economics and Business&2008& A < A+\\
     ABS10 &ABS Academic Journal Quality Guide&Association of Business Schools& 2010&1 < 2 < 3 < 4 < 4*\\
       Den11 & Danish Ministry Journal list&Danish Ministry &2011&1 < 2\\
       HEC11 & HEC Paris Ranking&Hautes Etudes Commerciales de Paris&2011& C < B < B+ < A\\
    UQ11 & Adjusted ERA Rankings List&University of Queensland&2011& 4 < 3 < 2 < 1\\
      Vhb11 & VHB ranking&Association of University Professors of Business in German speaking countries& 2011 &E < D < C < B < A < A+ \\
      Aeres12 & AERES journal list&Agence d'evaluation de la recherche et de l'enseignement superieur& 2012&C < B < A\\
       Cra12 & CRA ranking&Cranfield University School of Management&2012& 1 < 2 < 3 < 4\\
    EJL12 &ERIM Journals Listing& Erasmus Research Institute of Management&2012&S < P A < P < STAR\\
    Abdc13 & ABDC Journal Rankings & Australian Business Deans Council  &2013& C < B < A < A*\\
    Cnrs13 &CNRS ranking& Centre National de la Recherche Scientifique & 2013& 4 < 3 < 2 < 1 < 1*\\
    Ess13 & ESSEC journal ranking&ESSEC Business School Paris &2013 & 3 < 2 < 1 < 0 < 0+\\
     Fneg13 & FNEGE ranking& French Management Association & 2013&4 < 3 < 2 < 1 < 1*\\
 \label{tab:ranking}
  \end{longtable}
\begin{minipage}{\linewidth}
 \renewcommand{\footnoterule}{}
 \footnotetext{$^1$ based on Web of Science\textsuperscript{\textregistered} database}
  \footnotetext{$^2$ based on Scopus\textsuperscript{\textregistered} database}
\end{minipage} 
\end{scriptsize} 
\end{landscape}

\end{appendices}

\end{document}